# Music and musical sonification for the rehabilitation of Parkinsonian dysgraphia: Conceptual framework

Lauriane Véron-Delor[1,2], Serge Pinto[2], Alexandre Eusebio[3], Jean-Luc Velay[1,2], Jérémy Danna[1,2],

[1] Aix Marseille Univ, CNRS, LNC, Marseille, France
jean-luc.velay@univ-amu.fr; jeremy.danna@univ-amu.fr
[2] Aix Marseille Univ, CNRS, LPL, Aix-en-Provence, France
lauriane.veron-delor@univ-amu.fr; serge.pinto@univ-amu.fr
[3] Aix Marseille Univ, CNRS, INT, Marseille, France
alexandre.eusebio@univ-amu.fr

**Abstract.** Music has been shown to enhance motor control in patients with Parkinson's disease (PD). Notably, musical rhythm is perceived as an external auditory cue that helps PD patients to better control movements. The rationale of such effects is that motor control based on auditory guidance would activate a compensatory brain network that minimizes the recruitment of the defective pathway involving the basal ganglia. Would associating music to movement improve its perception and control in PD? Musical sonification consists in modifying in real-time the playback of a preselected music according to some movement parameters. The validation of such a method is underway for handwriting in PD patients. When confirmed, this study will strengthen the clinical interest of musical sonification in motor control and (re)learning in PD.

# 1 Introduction: external cueing and feedback as part of rehabilitation in Parkinson's disease

Parkinson's disease (PD) is the second most common neurodegenerative disorder after Alzheimer's disease. It is caused by the loss of dopaminergic neurons in the *pars compacta* of the *substantia nigra* and other neurological systems, leading to a set of motor and non-motor symptoms [1]. PD symptoms are managed with medication (e.g., L-Dopa, dopaminergic agonists) and/or neurosurgical interventions, including mainly deep brain stimulation. Nevertheless, limitations of such treatments to relieve motor disturbances have led to investigate non-pharmacological additional methods based on assisted motor rehabilitation. Among the various methods of motor rehabilitation, there is a growing interest in applying external cues and/or supplementary feedback to supplement drugs-based approaches.

Gait (for reviews, see [2] and [3]), and, to a lesser extent, handwriting (for a review, see [4]) were particularly brought into focus. The present chapter aims at reporting and questioning the studies carried out in the last decade and for which the effect of rehabilitation based on feedback or on auditory cueing was evaluated in Parkinsonian walking or handwriting (see Table 1).

**Table 1.** Auditory cueing- or feedback-based studies on gait, tapping, and writing in PD.

| | | **Auditory cueing-based rehabilitation studies** | | |
|---|---|---|---|---|
| Ref | Subjects | Conditions | Data analysis | Main results |
| [5] | 15 PD ON & 20 CTL | *12 trainings* Individualised RAS at 3 tempos, embedded in a musical structure | BAASTA; Stride length | Improvement in synchronization and hand tapping after training |
| [6] | 22 PD ON in 2 groups | *39 trainings* Individualised music vs. no music | Gait velocity; Stride time; Stride length; Cadence | Improvement of gait velocity, stride time and cadence following music training |
| [7] | 12 PD OFF | *Two tasks: walking then walking + carrying a cup full of water, under 4 conditions:* No cue vs. Visual (transverse strips) vs. Auditory (metronome) vs. visual and auditory cues | Freezing number and duration; Cadence; Gait velocity; Stride length | Improvement of cadence and stride length with visual and dual cues in both tasks Improvement of FOG with all types of cues in both tasks |
| [8] | 14 PD ON & 20 CTL | *12 trainings* - Individualised RAS at 2 tempos, embedded in a musical structure | BAASTA; Gait velocity; Stride length; Cadence; Synchronization variability | Improvement of PD gait parameters (velocity and stride length) directly and 1 month after training |
| [9] | 15 PD ON | *Two tasks (digital tapping + foot tapping) under 2 conditions:* No cue vs. auditory cue (metronome) | Freezing duration; Tapping frequency; Tapping amplitude | Metronome decreased the frequency and the incidence of freezing, and improved both digital and foot tapping frequency |
| [10] | 58 PD ON in 3 groups | *60 dance lessons* Tango vs. Waltz/foxtrot vs. Nothing | Balance; Gait velocity; Forward and backward walking; | Improvement in balance, gait velocity and backward walking in both dance groups Greater improvement with tango |
| [11] | 75 PD ON in 4 groups | *40 dance lessons* Tango vs. Waltz/foxtrot vs. Tai Chi vs. Nothing | HRQoL | Improvement in HRQoL only after tango |
| [12] | 20 PD ON in 2 groups | *3 trainings* SDTT vs. RAC | Gait velocity; Cadence; | RAC improved gait speed and SDTT improved balance |

| | | | Balance; HRQoL | Retention effects founded 3-month after both RAC and SDTT training |
|---|---|---|---|---|
| [13] | 47 PD ON in 2 groups | *8 days training* RAS vs. No cue | FOG number; Gait velocity; Stride length | Improvement of all gait parameters after RAS training |
| [14] | 25 PD ON (with vs. without FOG) & 10 CTL | *Walking session under 3 conditions*: Visual (transverse strips) vs. Auditory (metronome) vs. No cue | FOG number; Step number | Improvement in gait and FOG number with visual cue in PD FOG only No effect of auditory cue in PD FOG Better improvement in gait with auditory cues for PD without FOG than for PD FOG |
| [15] | 10 PD OFF | *Walking session under 2 conditions*: No cue vs. RAS (metronome) | Gait velocity; Stride length; Cadence | Improvement in all gait parameters after RAS training |
| [16] | 9 PD ON | *Walking session under 3 conditions*: No cue vs. CUET vs. CUEST | Gait velocity and variability; Stride amplitude; Cadence | Improvement of gait velocity, stride amplitude and cadence with both CUET and CUEST, the latter being the most effective |
| [17] | 10 PD ON & 10 CTL | *3 walking sessions: Session 1 under 4 conditions*: Verbal instruction vs. verbal instruction + metronome vs. HFGS vs. HFGS + verbal instruction *Session 2 under 4 conditions*: HFGS vs. HFGS + verbal instruction vs. synthesized footstep sounds vs. synthesized footstep sounds + verbal instruction *Session 3 under 4 conditions*: HFGS vs. mental imagery of HFGS vs. synthesized footstep sounds vs. mental imagery of synthesized footstep sounds | Stride length and variability; Velocity; Cadence; Gait variability | Decrease of stride length variability in PD patients during session with HFGS and HFGS + verbal instruction Improvement in stride length in all conditions except in synthesized sounds condition PD patients fail to adapt to the synthesized footstep sounds Performances are better during cueing than during imagery Performances are better during mental imagery of HFGS than during mental imagery of synthesized footstep sounds |
| [18] | 19 PD OFF (with vs. | *Walking session under 4 conditions*: Healthy footstep | Step time variability; | No cueing effect in PD without FOG |
| | without FOG) | on a corridor sounds vs. metronome vs. healthy footstep on gravel sounds vs. synthesized footstep sound | Swing time variability; Rhythmicity; Asymmetry | Improvement in temporal regularity in PD with FOG in forth conditions |
| **Auditory feedback-based rehabilitation studies** | | | | |
| [19] | 16 PD ON | *Walking session:* Clicking sound in response to every step | Cadence; stride length | Improvement in speed and stride length during and after training with FB |
| [20] | 42 PD ON (with vs. without FB) | *20 trainings with visual movement FB, visual color target FB and auditory target FB* | Clinical motor evaluations | Improvement in balance during and 1 month after experimental training No change in without FB group |
| **Auditory cueing- and feedback-based rehabilitation studies** | | | | |
| [21] | 28 PD ON (with vs. without FOG) | *6 weeks trainings under 4 conditions:* RAS vs. IC vs. IF vs. No cue/FB | Gait deviations | Gait deviations decrease with RAS in PD with FOG |
| [22] | 11 PD ON & 11 CTL & 11 CTL young | *Walking session under 4 conditions:* no cue vs. auditory cue vs. verbal instruction vs. COM (auditory cue and verbal instruction) | Gait velocity; Stride length; Cadence | Improvement of gait velocity and stride length with verbal feedback and COM |
| [23] | 15 PD ON & 15 CTL & 15 CTL young | *Unimanual and bimanual drawing sessions under 3 conditions:* visual cue vs. auditory FB vs. verbal FB | Amplitude; Amplitude variability; Coordination; Precision | Improvement of coordination and amplitude variability with both auditory and verbal feedback |
| [24] | 206 PD ON | *4 weeks treadmill training with visual and auditory FB and cues* | Steps length; Cadence; Coefficient of variance of both steps | Improvement of step length and variability, and cadence |

Abbreviations: PD: Individuals with PD; ON: on-medication; OFF: off-medication; CTL: Control subject; RAS: Rhythmic auditory stimulation; BAASTA: Battery for the assessment of auditory sensorimotor and timing abilities, including timing perception, discrimination and synchronization; HRQoL: Health related quality of life; SDTT: Speed-dependent treadmill training; RAC: Rhythmic auditory cue, individualised music playlist and metronome; FOG: Freezing of gait; CUET: Cue temporal, metronome with temporal instruction ''As you walk try to step in time to the beat''; CUEST: Cue spatiotemporal, metronome with spatiotemporal instruction ''As you walk try to take a big step in time to the beat''; HFGS: Healthy footstep on gravel sounds; IC: Intelligent cueing (auditory rhythm signal when strides deviated more than 5% from the reference cadence); IF Intelligent feedback (verbal instruction to speeding or slowing); FB: Feedback; COM: Combined information auditory cue and verbal instruction.

The organization and production of movement involves the integration of sensory information, which can be considered as basic feedback. Basic feedback informs about both the environment and the current state of the body to determine the appropriate set of muscle forces to generate the desired movement. Thus, a deficit in the processing of basic feedback affects the initiation of movements. Vision and/or audition are the most commonly used sensory modalities to support initiation and control of movement. They are differentially specialized to encode information from the environment and our body, visual information being more relevant for spatial processing, and auditory information for temporal processing [25].

On the one hand, supplementary feedback enriches the perception of self-performance during or after movement production, mainly based on internal expectations/representations. Supplementary feedback can provide information about the outcomes of an action with respect to the environmental goal or about the process, i.e. the movement produced. The terms of "knowledge of results" and "knowledge of performance" are respectively employed [26]. On the other hand, external cues yield a point of reference to guide movements execution [27]. In the field of motor rehabilitation with PD patients, the use of auditory cues has been largely preferred, especially for improving gait (e.g., [2] and [28]) and speech (e.g., [29] and [30]) disorders. Such enthusiasm is certainly justified by the natural and spontaneous tendency in humans to synchronize action with rhythm [31]. Very promising, and sometime unexpected, effects have been observed with the use of rhythmic auditory stimulation (RAS; see table 1 – e.g., [8]). PD patients are tempted to couple their steps to RAS provided by a metronome or an amplified beat of a music. Some researches reveal improvements with RAS in gait velocity and stride length, sometimes with long-term benefits [8] and [13]. Music itself carries an intrinsic rhythm that plays the role of an external cue (e.g., [32]), as a metronome, guiding movements. Moreover, music contributes to something more than simple metronome rhythm: emotional aspects are conveyed with the melody, especially when the music is familiar. Music involves both cognitive and emotional processing, which can be used to carry over effects (e.g., mental singing – [33]). In healthy individuals, Wittwer and colleagues [34] have compared effects of rhythmic music and metronome as external cuing on gait. They showed that music might be more efficient than the metronome to improve the velocity and cadence of gait, due to emotional aspects and motivation ensuing by melody. In individuals with PD, it has been shown that continuous sounds, like music, lead to better gait fluency than a simple metronome (see table 1, [18]).

## 2   Why does supplementary feedback or external cueing facilitate motor (re)learning for individuals with PD?

Movement rehabilitation in PD aims at improving motor control and coordination by either strengthening pre-existing pathways [27] or creating alternative circuits bypassing the basal ganglia. Motor learning is possible in individuals with PD (for a review, see [4]). Such concept raises the question of brain neuroplasticity. Neuroplasticity encompasses the ability for healthy neural networks to form new synapses in order to bypass and reorganize the damaged network [27]. Any functional motor rehabilitation is based on this phenomenon and may facilitate neurological recovery [35]. Neuroplasticity is stimulated by frequent motor or cognitive activities. Nevertheless, it is slowed down in individuals with PD compared to healthy subjects. This must be taken into account in the rehabilitation duration [4].

Distinct phases, consolidation, automatization, and retention, are identified in the process of motor learning. Doyon & Benali [36] revisited a model describing the brain plasticity during motor learning. According to this model, a clear distinction is proposed between motor sequence learning (MSL), which characterizes the process by which practice turns a sequence of actions into a behaviour, and motor adaptation (MA), which is required in response to environmental changes. Motor learning goes along with a decrease of cortical activity, especially in prefrontal and parietal regions that are involved in attentional processing of sensory information. At the same time, the activation of the cerebellum and basal ganglia increase according to the type of motor task, MA and MSL, respectively [37].

PD affects the functioning of the striato-thalamo-cortical loop (Figure 1, dotted arrows), particularly involved in the control of learned movements. Two possibilities can be proposed: On the one hand, the injured network (Figure 1, black arrows) could be restored similarly as pharmacological treatments [38]. On the other hand, a compensatory neural mechanism could be used to bypass the damaged pathway: The cerebello-thalamo-cortical loop (Figure 1, grey arrows) is involved in the control of movements in MA and seems preserved in PD, at least in the early stages of PD [4] and [39].

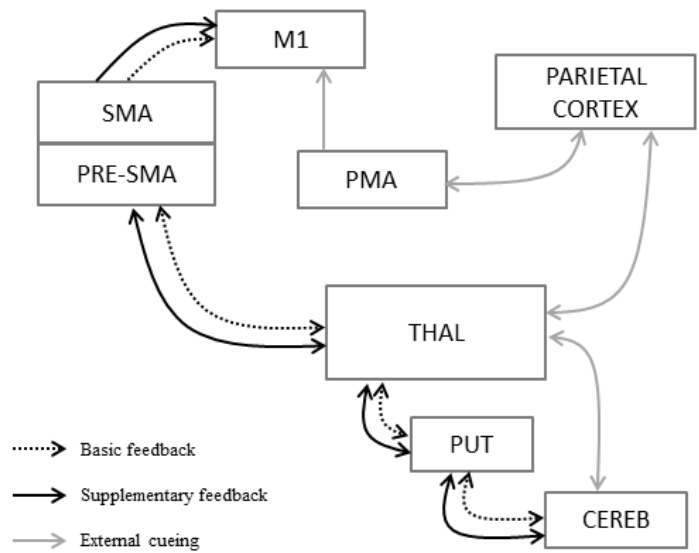

**Figure 1.** Motor control loops: two options to restore efficient motor control in PD patients (adapted from [4] and [39]). SMA: Supplementary motor area; PRE-SMA: Pre-supplementary motor area; M1: Primary motor area; PMA: Premotor area; THAL: Thalamus; PUT: Putamen (one of the basal ganglia); CEREB: Cerebellum.

**2.1 Applying supplementary feedback in PD patients "to restore the pathway"**

A deficit of sensory integration in PD has been documented by several studies [40], [41] and [42]. Regarding vision, it has been shown that visual withdrawal leads individuals with PD to increase their movement amplitude [43] and to reduce their velocity [44]. Such effects were not observed in healthy subjects. Longstaff and colleagues [45] proposed that moving slower would be a strategy of PD patients to improve online control, i.e. to be more feedback-dependent [42]. In healthy subjects, the absence of visual feedback can be compensated by kinaesthetic feedback. In PD patients, kinaesthetic feedback does not inform correctly about the hand or upper limb position and movement [42]. Therefore, the absence of visual feedback cannot be fully compensated in PD patients.

Beyond informational processing, applying supplementary feedback in a learning or rehabilitation protocol affects motivation. For example, providing learners with feedback after correct trials, compared with after incorrect trials, results in more effective learning [46]. Interestingly, basal ganglia are critical for supporting learning that is driven by feedback and is motivated by rewards [47]. Foerde and Shohamy [48] reported that the midbrain dopamine system supports feedback-dependent learning processes essential for predicting outcomes. Therefore, applying a real-time supplementary feedback would be relevant for restoring the reward network in PD patients.

**2.2 Applying external cueing in PD patients "to hit another pathway"**

Motor control and coordination are managed by both basal ganglia and cerebellum. Thus, promoting the cerebellum activation to bypass basal ganglia appears as a relevant strategy of rehabilitation. Nombela and colleagues [39] have gathered the findings of various neuroimaging studies in which the auditory external cueing on PD gait was evaluated. Their review provides an accurate description of how music influences motor mechanisms. RAS in music or metronome can act as an external "timer" guiding the execution of the movement and bypassing the dysfunction in striato-thalamo-cortical loop [39] and [49]. When the movement is performed with an external cueing, the online control of movements becomes dependent to this supplementary environmental constraint: the task tends to become similar to a MA task.

## 3  Effects on Parkinsonian dysgraphia

Handwriting is a complex motor activity that requires a great level of expertise. Interestingly, handwriting is particularly altered by PD [50], [51] and [52]. Handwriting disorders in PD are mainly known from the observation of an abnormal reduction in writing size so-called micrographia [53]. Micrographia affects about 50% of individuals with PD. According to Van Gemmert et al. [54], micrographia would result from an inability to maintain a constant force during handwriting, as well as to synchronize wrist and finger movements. Consequently, beyond micrographia, other kinematic and dynamic variables (velocity, dysfluency, i.e. abnormal velocity fluctuations, etc.) would be more systematically altered in Parkinsonian handwriting. Therefore, the term *Parkinsonian dysgraphia* has been proposed [55] and [56].

What are the causes of PD dysgraphia? On the basis of different models of handwriting, such as the kinematic model [57] and [58] or the neural model of handwriting [59], the "stroke" – the basic motor unit of handwriting – results from the coordinated activity of the muscular system coded as a velocity vector. Interestingly, in these models, only the orientation and amplitude of each velocity vector is processed in the central nervous system and this process is precisely achieved in basal ganglia that are affected in PD (e.g., [60]). Another argument concerns the nature of the task that changes in the course of learning. In beginners, handwriting is like a MA task: they must correct the ongoing movements of the pen thanks to the visual inspection of the generated written trace. Once the characters are learned, the underlying motor pattern is automatized, and handwriting becomes mainly a sequential task in which the writers must check the very rapid succession of the strokes composing a character and the correct sequences of characters composing a word. According to Doyon & Benali's model [36], this transition relative to the nature of the task would be associated with a switch from the cortico-cerebellar loop, more activated at the early stage of learning, to the cortico-striatal loop, more activated at the latest stage. This assumption was investigated and partially validated in a combined fMRI and kinematic study conducted in healthy adults during a fast-learning of a graphomotor sequence [61]. If confirmed, this may explain both why handwriting is altered in individuals with PD, and why external cueing or supplementary feedback may be particularly relevant for helping them to better control their handwriting.

### 3.1 The classical method of handwriting rehabilitation for PD patients

In 1972, McLennan [53] suggested that the mere presence of parallel lines could allow individuals with PD to maintain their writing size, thus improving micrographia. This method was tested and validated several times in graphomotor tasks [23] and [62]. Other visual cues, as target points [63] or grid lines [64], has been tested and the authors have shown that they improve both the writing size and width. Furthermore, these cues allow the patients to maintain a correct size during the entire task [62] and [64]. Another method was tested with a graphic tablet [65]. The written trace was displayed in real-time on a screen in front of the writers and their hand and pen were hidden in such a way that participants had visual feedback about the written trace only. This feedback was either normal, smaller, or larger than the actual handwriting. The authors observed that individuals with PD succeed in the visuo-motor adaptation by changing the amplitude of their writing movement when the visual feedback was distorted. However, such effects were present when the hand was hidden only and disappeared when the hand was not hidden [66]. Beyond improvements of the spatial feature of handwriting, Nieuwboer and colleagues also demonstrated that freezing of upper limb was improved by visual cueing in a drawing task [62].

When comparing the effects of visual cueing and auditory feedback individuals with PD performed better in a graphomotor task when they received an auditory feedback based on verbal instructions or on a spatial sonification than when they realized the task with the presence of visual cues solely [23]. However, we cannot conclude whether the advantage of applying auditory feedback rather than visual cueing results from the use of feedback, the auditory modality, or both. Note that the positive effect of auditory cueing was not observed in a bimanual drawing task by Swinnen and colleagues [67].

### 3.2 Towards a new method of handwriting rehabilitation with PD patients based on musical sonification

The presence of auditory feedback or cueing improves significantly the motor control of individuals with PD. On the one hand, providing a supplementary auditory feedback enriches the patients' perception of their movements and thus enhances their control. On the other hand, providing an external auditory cueing leads the patient to adapt their movements in a very promising way. Is it possible to combine the advantages of both methods?

In this international symposium on computer music multidisciplinary research (CMMR 2017), an individualized approach in the use of RAS was proposed to help PD patients to walk [8]. The principle was to adapt the RAS in real-time to patients step times. The results revealed important individual differences among PD patients with regard to their response to different cueing strategies. The strategy that we are currently evaluating differs from that: we are assessing the effect of abrupt changes of music linked to kinematic thresholds. This method of m*usical sonification* consists of modifying a preselected music according to movement variables: music is distorted when the movement is dysfluent and too slow. The aim is both to improve the perception of movement irregularities (when music changes) and to provide an auditory guidance (when music does not change).

The melodious music associated to a correct movement supplies the writer with an auditory cueing based on musical rhythm. Moreover, melody is also a reward motivating the patients, provided that it is pleasant. This strategy of musical sonification allows patients:

a) To use music as an external cue, considering the advantages of musical rhythm and RAS effects on motor control in PD patients that we have previously described (e.g. [31] and [35])

b) To use music as an auditory feedback informing about the movement correctness if s/he has some difficulties in synchronizing his/her movements with the musical rhythm. Indeed, the ability to synchronize movements with an external rhythm requires a temporal processing on both the metronome and the movement itself. This concomitant processing is potentially affected in PD because it involves the cortico-striatal loop [8]. In the present strategy, writing becomes a pseudo-musical practice. The presence of kinematic thresholds, which can be individualized, leads the writer to manage music with the pen like an orchestra conductor with the baton. The writer can stop and start music when s/he decides to. The pen shapes and reshapes music. Consequently, music can be considered as an external goal on which the patients pay attention. Actually, it has been shown that the external focus of attention enhances motor control in PD patients (e.g., [68]).

c) To integrate music as auditory feedback *and* auditory guidance if they succeed in taking advantage of both supports.

We are currently evaluating this strategy of musical sonification with PD patients and healthy controls. The experiment is designed as a "pre-test/training/post-test" with three different training sessions: one with music, one with musical sonification, and one in silence. During the tests (all in silence), participants were asked to draw loops, write the French word "*cellule*" (cell) in cursive, and make their own signature (for an illustration, see figure 2).

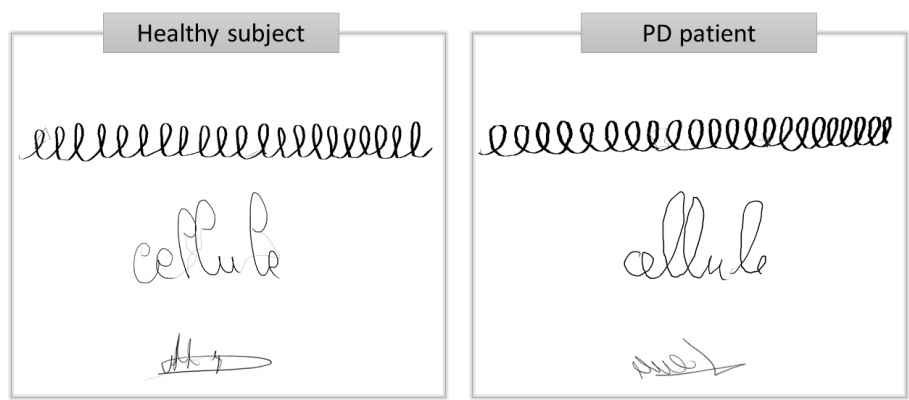

**Figure 2.** Handwriting tasks produced during the pre-test by a healthy subject (left) and an individual with PD (right).

During the training phases, the participants were required to achieve graphomotor exercises under one of the three conditions (music vs. musical sonification vs. silence). The conditions order was counterbalanced between participants and all participants were tested just before and after each training. When the performance of both groups during each training session (in silence, music or musical sonification) were compared, the very preliminary results (on nine PD patients and nine controls) revealed that writing speed was much higher in both groups under musical sonification. When the differences of performance between post- and pre-test were compared for each training session, both PD patients and controls were faster after musical sonification, both in drawing loops and word writing. These preliminary findings must be interpreted with many cautions. If confirmed, they show that PD patients better perform the task under musical sonification and maintain these improvements at short term. Therefore, musical sonification would be a very promising rehabilitation method for individuals with PD.

## 4   Conclusions

In the digital age, the interest of handwriting rehabilitation in PD may be limited, although writing a short message on a sticky note or a shopping list is still very useful in the daily life. The advantage for the patient lies rather in the possible transfer of the effects to fine motor rehabilitation. Beyond handwriting, the rehabilitation of the "clumsy hands" that hampers the activities of daily life [69] significantly improves the patients' quality of life, in eating, getting dressed, washing *etc.* [23] and [27]. Furthermore, motor rehabilitation also slows down the degenerative processes related to PD. The positive effects of external cueing seem to persist over time as if it remains present "inside the head" whereas it is not physically present [4], similarly as basic auditory feedback when playing piano or when producing other audible motor activities [17] and [70]. However, a definite conclusion will be reached when the neural changes underlying the motor improvements following a rehabilitation based on musical sonification will be observed.

**Acknowledgments.** This research was supported by Grants ANR-16-CONV-0002 (ILCB), ANR-11-LABX-0036 (BLRI), and ANR-11-IDEX-0001-02 (AMIDEX) the Excellence Initiative of Aix-Marseille University. We want to thank Richard Kronland-Martinet, Solvi Ystad et Mitsuko Aramaki (laboratory PRISM), as well as Charles Gondre for the technical development related to the musical sonification.